\documentclass[final,3p,times]{elsarticle}
\usepackage{graphicx}
\usepackage{bm}
\newcommand{\ul}[1]{\underline{#1}}
\newcommand{\dif}[2]{\partial #1/\partial #2}
\newcommand{\diff}[2]{\frac{\partial #1}{\partial #2}}
\newcommand{\mapright}[1]{\smash{\mathop{\hbox to 1.5cm{\rightarrowfill}}\limits^{#1}}}

\journal{Physica A}

\begin{document}

\begin{frontmatter}

\title{The effects of the chemical potential in a BE distribution and the fractional parameter in a distribution with Mittag--Leffler function}

\author[1]{Minoru Biyajima}
\address[1]{Department of Physics, Shinshu University, Matsumoto 390-8621, Japan}
\ead{biyajima@azusa.shinshu-u.ac.jp}
\author[2]{Takuya Mizoguchi}
\address[2]{National Institute of Technology, Toba College, Toba 517-8501, Japan\corref{telephone/fax: +81599258088}}
\ead{mizoguti@toba-cmt.ac.jp}
\author[3]{Naomichi Suzuki}
\address[3]{Matsumoto University, Matsumoto 390-1295, Japan}
\ead{suzuki@matsu.ac.jp}

\begin{abstract}
The fractional Planck distribution is calculated by applying the Caputo fractional derivative with order $p$ ($p > 0$) to the equation proposed by Planck in 1900. In addition, the integral representation of the Mittag--Leffler function is employed to obtain a new formula for the fractional BE distribution, which is then used to analyze the NASA COBE monopole data. Based on this analysis, an identity $p\simeq e^{-\mu}$ is found, where $\mu$ is the dimensionless constant chemical potential that was introduced to the BE distribution by the NASA COBE collaboration.
\end{abstract}


\end{frontmatter}

\section{\label{sec1}Introduction}
There has been recent progress in physical sciences studies \cite{Metzler2000,West2003,Hilfer2000,Barkai2001,Suzuki2001,Biya2015} that are based on fractional calculus \cite{Caputo1967,Ross1975,Podlubny1999,Mainardi2001,Dzhe1993,Gore2002,Gore2014}. As part of these contributions, we investigate the NASA COBE monopole data \cite{Mather1994,cobe2005,Durrer2008} by utilizing the Bose--Einstein (BE) distribution and a fractional calculus based distribution \cite{Ertik2009,Biya2012,Biya2015}. A well-known solution for Kompaneets equation \cite{Komp1957,Weym1965}, which describes the photons distribution in the early Universe, is given by:
\begin{eqnarray}
  U^{\rm (BE)}(T,\,\nu,\,\mu) &=& \frac{C_B}{e^{x+\mu} - 1} 
   \mapright{\ |\mu| \ll 1\ }\ C_B\left[\frac 1{e^x-1} - \mu\frac{e^x}{(e^x-1)^2}\right]\nonumber\\
  &=& F^{\rm (PD)} + F_2^{\rm (BE)}(x,\ \mu),
\label{eq1}
\end{eqnarray}
where $x=h\nu/k_BT$ and $C_B = 2h\nu^3/c^2$. $\mu$ is a dimensionless chemical potential. See Table \ref{tab1} I) Kompaneets equation.

On the other hand, the fractional calculus based photons distribution of the Universe is given by (See Table \ref{tab1} II) Fractional calculus and III) Planck distribution in 1900 \cite{Planck1900a,Planck1900b,Sommerfeld1956}):
\begin{eqnarray}
  U(x)  = \frac{C_B}{ E_p(x^p) - 1},
\label{eq2}
\end{eqnarray}
where $p$ is the fractional parameter. $E_p(x^p)$ is the Mittag--Leffler (ML) function defined as  
\begin{eqnarray}
  E_p(x^p) = \sum_{n=0}^{\infty} \frac{x^{np}}{\Gamma(np + 1)}.
\label{eq3}
\end{eqnarray}
The following digamma function $\psi$ is used in ref. \cite{Biya2015} to analyze the NASA COBE data,
\begin{eqnarray}
  \mbox{Eq.\ }(\ref{eq2})\ \mapright{\ |p-1| \ll 1\ } &\!\!\!&\!\!\!
  C_B\left[\frac 1{e^x-1} + \frac{p-1}{(e^x-1)^2}\right. 
   \left. \sum_{k=0}^{\infty} \frac{kx^k[\psi(k+1)-\ln x]}{\Gamma(k+1)}\right]\nonumber\\
  &\!\!\!&\!\!\! = F^{\rm (PD)} + F_2^{\rm (FC)}(x,\ p-1),
\label{eq4}
\end{eqnarray}
where $\psi(z) = d(\ln \Gamma (z))/dz$.

The COBE data is analyzed using Eqs. (\ref{eq1}) and (\ref{eq2}), thereby yielding the following estimated values:
\begin{eqnarray}
  &&T = 1/k_B\beta = 2.725 \mbox{ K},\ |\mu| < 7.58\times 10^{-5},\nonumber\\
  &&|p-1| < 8.09\!\times\! 10^{-5},\ |\mu|\!\times\! I_1 < 5.47\!\times\! 10^{-4}\ (1.00),\nonumber\\
  &&|p-1|\times I_2 < 5.60\times 10^{-4}\ (1.03),
\label{eq5}
\end{eqnarray}
where, with Riemann's $\zeta$ function $\zeta(3)$,
\begin{eqnarray*}
  I_1 &=& \int_0^{\infty}\frac{x^3\cdot e^x}{(e^x-1)^2}dx = 3 \cdot 2 \cdot \zeta(3),\\
  I_2 &=& \int_0^{\infty}\frac{x^3}{(e^x-1)^2}\sum_{k=0}^{\infty} \frac{kx^k[\psi(k+1)-\ln x]}{\Gamma(k+1)}dx\\
      &=& \sum_{k=0}^{\infty}\sum_{m=2}^{\infty} \frac{k(m-1)}{\Gamma(k+1)}\left(\frac 1m\right)^{k+4}\Gamma(k+4) 
  \left[\ln m - \left(\frac 1{k+3}+\frac 1{k+2}+\frac 1{k+1}\right)\right].
\end{eqnarray*}
We would like to pay our attention to these similar values, $\mu$ and $p-1$. The ratio between $|\mu|\times I_1$ and $|p-1|\times I_2$ is 1.00 to 1.03.

The same analysis of the COBE data is presented in this study, where we would like to adopt an integral representation of the ML function:
\begin{eqnarray}
  E_p(x^p) = \frac{e^x}p + \delta(p,\ x), 
\label{eq6}
\end{eqnarray}
where
\begin{eqnarray} 
  \delta(p,\ x) = -\frac{\sin(p\pi)}{\pi} \int_{0}^{+\infty} \frac{ y^{p-1} e^{-xy} }{ y^{2p}  - 2y^{p}\cos(p\pi) + 1 } dy.
\label{eq7}
\end{eqnarray}
A detailed derivation of the integral representation is supplied in \S \ref{sec2}. The magnitude of the integral representationi$\delta(x,\ p))$ contribution is estimated through concrete analysis of the COBE data in \S \ref{sec3}. Through such analysis of the COBE data, the following relation ensues:
\begin{eqnarray} 
  \mu = -\ln p\,.
\label{eq8}
\end{eqnarray}
The concluding remarks and discussion are provided in \S \ref{sec4}.
\begin{table}[htbp]
\centering
\caption{\label{tab1}Kompaneets equation, the fractional and the basic equations of the Planck distribution.}
\begin{tabular}{|l|}
\hline
{\bf I)} \ul{Kompaneets Eq. describing the photons distributions 
in the early Universe} \cite{Komp1957,Weym1965}\\
It is given by\\
\qquad $\dif ft =  C_{\kappa}x_e^{-2}\diff{}{x_e}x_e^4(\dif f{x_e} + f+ f^2)$\\
where $C_{\kappa} = (k_B T_e/m_ec^2)(n_e\sigma_e/c)$ and $x_e = h\nu/k_BT_e$. $T_e$, 
$T$, $n_e$, and $\sigma_e$ are the electron temperature,\\ radiation
temperature, electron density, and Thomson's cross-section,
respectively. As a stationary\\ solution, we derive the Bose
--Einstein (BE) distribution with the chemical potential,\\
\qquad $f(x) = 1/(ce^{x} - 1) = 1/(e^{x_e+\mu} - 1).$\\
\hline
{\bf II)} \ul{Fractional calculus} \cite{Caputo1967}\\
For the stationary solution in I) and a solution in III), if 
an inverse function $R(x)=1/f(x)$ exists,\\ the following 
equation is obtained,\\
\qquad $\dif Rx = R(x) + 1.$\\
The Caputo derivative in fractional calculus is applied to 
the previous equation, {\it i.e.},\\
\qquad ${}_0D^p_x R(x) = R(x) + 1,$\\
The Caputo fractional derivative of the $f(x)$ function for 
$m=1,2,\ldots$ is given by\\
\qquad ${}^C_0\! D^p_x f(x) = \frac{1}{\Gamma(m-p)} \int_0^x (x-\tau)^{m-p-1} f^{(m)}(\tau) d\tau,
\quad (m-1 < p < m)$,\\
and\\
\qquad $\displaystyle{ \lim_{p\rightarrow m}{}^C_0\! D^p_x f(x) = f^{(m)}(x)} = d^m f(x)/dx^m$.\\

The following distribution is calculated using the ML 
 function,\\
\qquad $f(x) = 1/R(x) = 1/(E_p(x^p) - 1).$\\
\hline
{\bf III)} \ul{Planck distribution in 1900} \cite{Planck1900a,Planck1900b,Sommerfeld1956,Tsallis2005}\\
Planck utilized the following equation to describe the 
photons distribution $U$,\\
\qquad $\dif U{\beta} = - U - U^2,$\\
Adopting the aforementioned method in section II), the 
same expression is deduced.\\
\hline
\end{tabular}
\end{table}

\section{\label{sec2}Integral representation of the Mittag--Leffler function}
The integral representation of the Mittag--Leffler (ML) function $E_\alpha(z)$~\cite{Dzhe1993, Podlubny1999, Gore2002, Gore2014} is given by,
\begin{eqnarray}
  &&E_{\alpha}(z) = -\frac{ 1 }{ 2\pi i \alpha} \int_{\gamma(\varepsilon;\delta)} 
  \frac{e^{ \zeta^{1/\alpha}}}{\zeta - z} d\zeta + \frac{1}{\alpha} e^{z^{1/\alpha}}, 
  \quad z\in G^{(+)}(\varepsilon;\delta), 
\label{eq9}
\end{eqnarray}
\begin{eqnarray}
  E_{\alpha}(z) = -\frac{ 1 }{ 2\pi i \alpha } \int_{\gamma(\varepsilon;\delta)} 
  \frac{ e^{ \zeta^{1/\alpha} } }{ \zeta - z } d\zeta, \quad
  z\in G^{(-)}(\varepsilon;\delta),\quad
\label{eq10}
\end{eqnarray}
in the complex plane, $z=x+iy$ ( $x= Re\, z$, $y= Im\, z$), under the conditions,
\begin{eqnarray}
  0 < \alpha <2, \quad  \pi\alpha/2 < \delta  \le  {\rm min}\{\pi,\pi\alpha\}.
\label{eq11}
\end{eqnarray}

\begin{figure}[htb]
  \centering
  \includegraphics[width=60mm,clip]{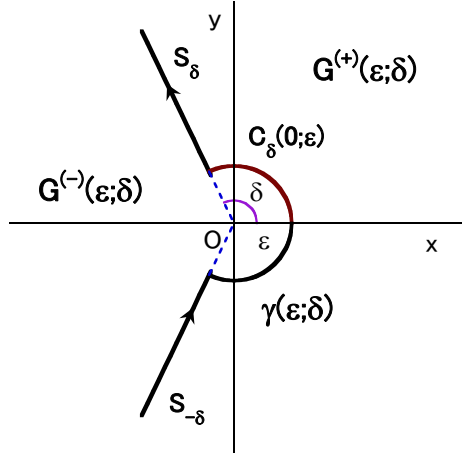} 
  \caption{\label{fig1} The contour $\gamma(\varepsilon;\delta)$ composed of the ray $S_{-\delta}$, circular arc $C_\delta(0;\varepsilon)$ and  ray $S_{\delta}$}
\end{figure}

As shown in Fig. \ref{fig1}, the contour $\gamma(\varepsilon;\delta)$ ($\varepsilon>0$, $0<\delta \le \pi$) is comprised of the following three parts: 
(i) ray $S_{-\delta}$ ( ${\rm arg}\,\, \tau = -\delta$, $|\tau| \ge \varepsilon$ ), 
(ii)  circular arc $C_\delta(0;\varepsilon)$ ( $-\delta \le {\rm arg}\,\, \tau \le \delta$, $|\tau| = \varepsilon$ ),  
(iii) ray $S_{\delta}$ ( ${\rm arg}\,\, \tau = \delta$, $|\tau| \ge \varepsilon$ ). The left hand side of the contour $\gamma(\varepsilon;\delta)$, where the origin O is included, and is denoted by $G^{(-)}(\varepsilon;\delta)$, while the other side is denoted by $G^{(+)}(\varepsilon;\delta)$D
 
After integrating along the contour $\gamma(\varepsilon;\delta)$ while taking the $\varepsilon \rightarrow 0$ limit into consideration, Eqs. (\ref{eq9}) and (\ref{eq10}) respectively reduce to,
\begin{eqnarray}
  &&E_{\alpha}(z) =  -\frac{z\sin(\pi\alpha)}{\pi \alpha} \int_{0}^{+\infty} 
  \frac{ e^{ -r^{1/\alpha} } }{ r^2 - 2rz\cos(\pi\alpha) + z^2 } dr 
  + \frac{1}{\alpha} e^{z^{1/\alpha}},\quad \in G^{(+)}(+0;\delta),
\label{eq12}
\end{eqnarray}
\begin{eqnarray}
  &&E_{\alpha}(z) = - \frac{z\sin(\pi\alpha)}{\pi \alpha} \int_{0}^{+\infty} 
  \frac{  e^{ -r^{1/\alpha} } }{ r^2 - 2rz\cos(\pi\alpha) + z^2 } dr,\quad 
  z\in G^{(-)}(+0;\delta).
\label{eq13}
\end{eqnarray}

If $0 <\alpha\le 1$,  the explicit expression for $E_\alpha(z)$ is obtained through Eqs.(\ref{eq12}) 
and (\ref{eq13}). If $\alpha>1$, using the ML function's summation formula, 
\begin{eqnarray}
  E_{\alpha}(z) = \frac{1}{m} \sum_{h=0}^{m-1} E_{\alpha/m} ( z^{1/m} e^{i2\pi h/m} ),
\label{eq14}
\end{eqnarray}
we can express  $E_\alpha(z)$ through those with suffixes $\alpha/m \le 1$. Then, the ML function's expression is given by the procedure in the case of $0 < \alpha < 1$.

\begin{figure}[htb]
  \centering
  \includegraphics [width=75mm,clip]{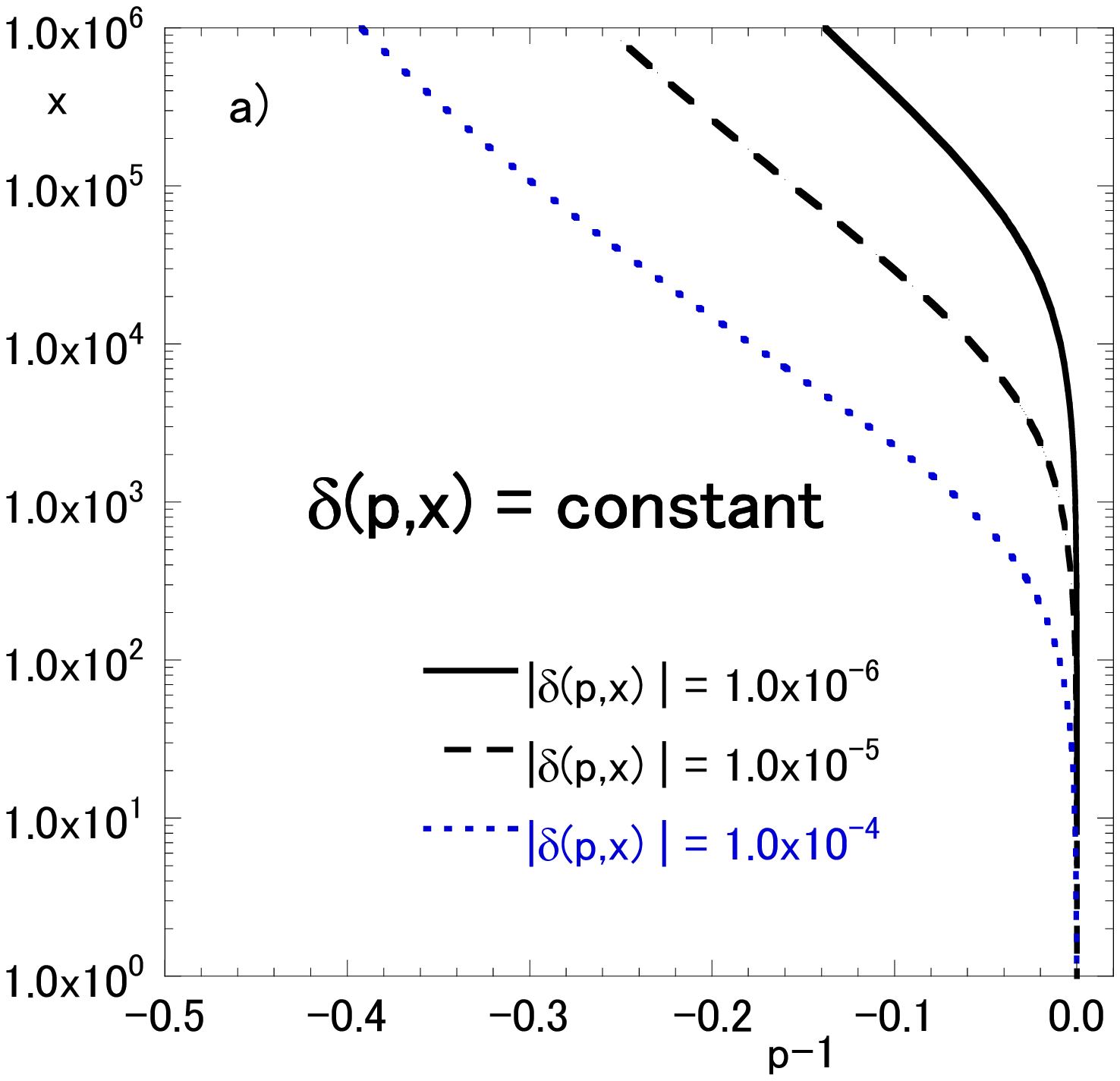}
  \hspace{2mm}
  \includegraphics [width=75mm,clip]{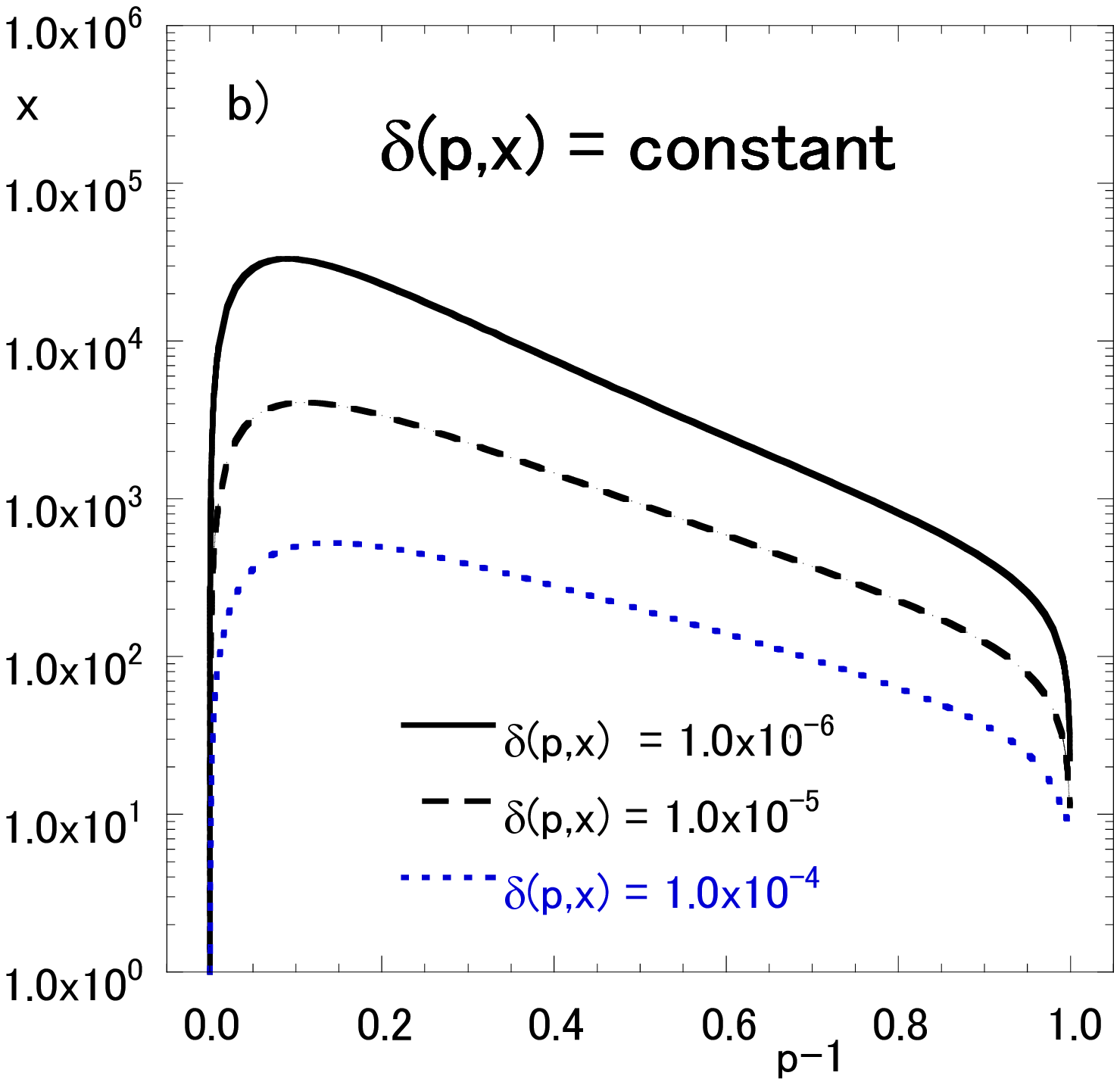} 
  \caption{\label{fig2} a) Contour maps of $\delta(p,\ x)=$ constant for $0.5<p<1$. b) Contour maps of $\delta(p,\ x)=$ constant for $1<p<2$.}
\end{figure}

In Fig. \ref{fig2}, we numerically examine the behavior of $\delta(p,\ x)$ in the ML function $E_p(x^p)$ for $0<p<2$ and $x \ge 0$. The integral representation $\delta(p,\ x)$ for $0<p$ defined in Eq.(\ref{eq7}) approaches to zero in the limit of $x \rightarrow \infty$. Function $\delta(p,\ x)$ at fixed $p$ for $0<p<1$ is negative and increases monotonously with $x$, and $\delta(p,\ x)$ for $1<p<2$ is positive and decreases monotonously with $x$. For $0 < p < 2$, the following limit can be derived through analytic calculations \cite{Gradshteyn1965}:
\begin{eqnarray}
  |\delta(p,\ x)| \le |\delta(p,\ 0)|  = |p-1|/p.
\label{eq15}
\end{eqnarray}

\section{\label{sec3}Analysis of COBE data by Eqs. (\ref{eq2}) and (\ref{eq6})}
Expanding Eq. (\ref{eq2}), providing that $(p-1)\ll 1$, as follows
\begin{eqnarray}
  U(T, \nu, p) &=& \frac{C_B}{e^x -1} + \frac{C_Be^x\ln p}{(e^x -1)^2} -\frac{C_B\delta(p,\ x)}{(e^x -1)^2}\nonumber\\
  &=& F^{\rm (PD)}(x) + F_2^{\rm (FC)}(x,\, \ln p) - F_3^{\rm (FC)}(x,\: \delta(p,\, x)),\quad
\label{eq16}
\end{eqnarray}
\begin{figure}[htb]
  \centering
  \includegraphics [width=80mm,clip]{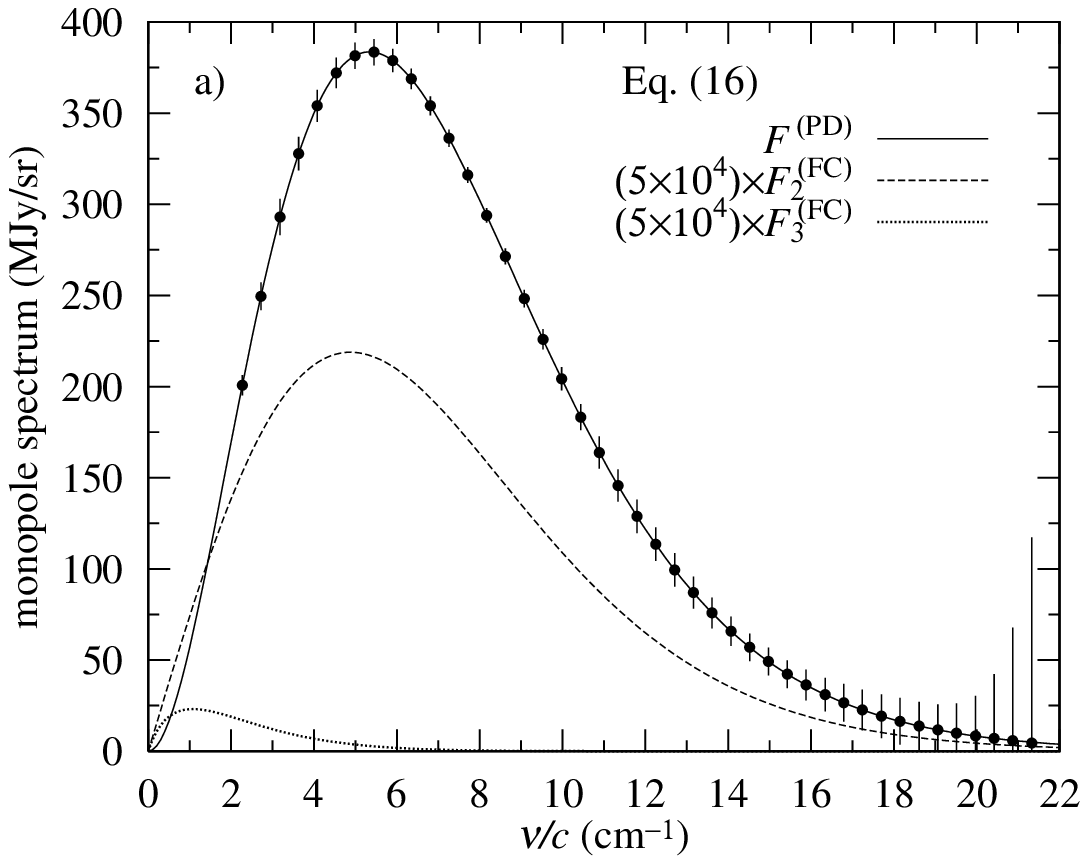}
  \includegraphics [width=80mm,clip]{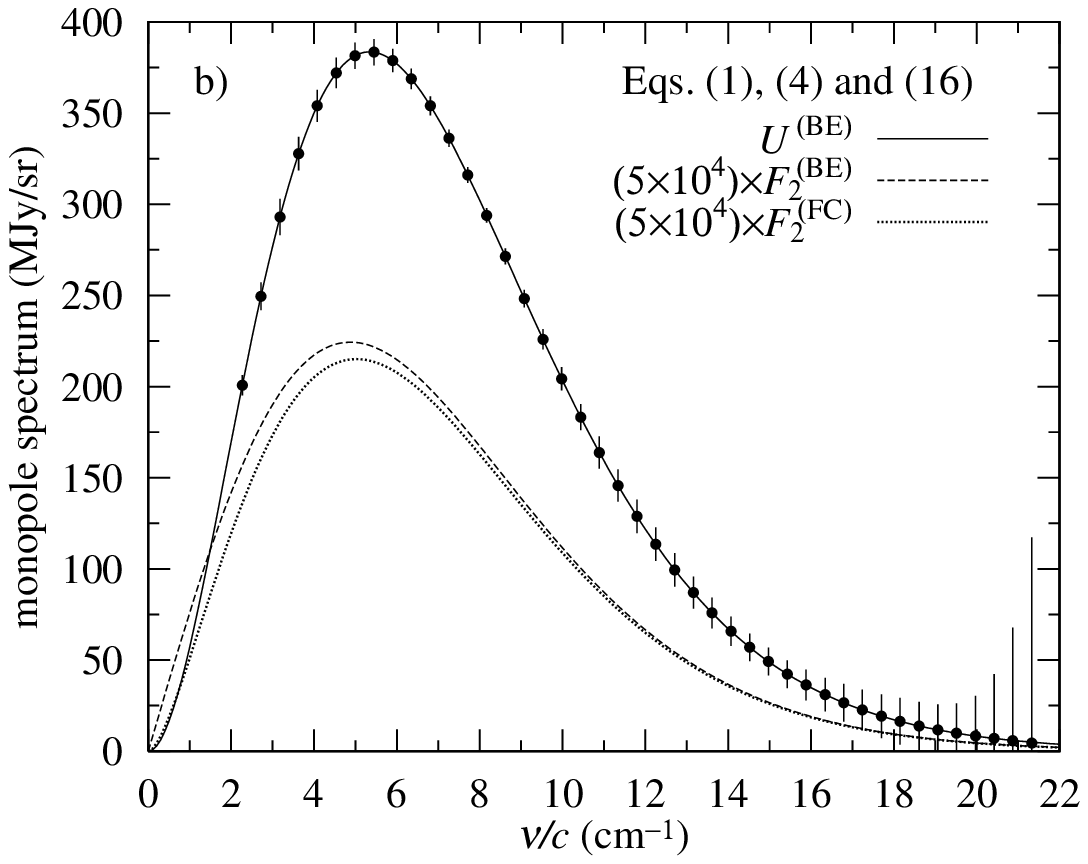}
  \caption{\label{fig3} Analysis of the COBE monopole data by Eqs.~(\ref{eq16}), (\ref{eq1}) and (\ref{eq4}). a) Eq.~(\ref{eq16}). b)  Eqs.~(\ref{eq1}), (\ref{eq4}) and (\ref{eq16}). The magnitudes of error bars mean that 400 times real error bars.}
\end{figure}
we can analyze the NASA COBE data in Fig. \ref{fig3}. For comparison, the results on $F_2^{\rm (FC)}(x,\ \ln p)$ and $F_3^{\rm (FC)}(x,\ \delta(p,\ x))$ are shown separately. As is seen in Fig. \ref{fig3}, it can be said that contribution from $F^{\rm (PD)}(x)$ is much bigger than that of $F_3^{\rm (FC)}(x,\ \delta(p,\ x))$. Then we can directly compare Eq. (\ref{eq1}) with $e^x/p$ and obtain Eq. (\ref{eq8}): $\mu = -\ln p$. From this expression, $p$ can be regarded as an inverse fugacity \cite{Pathria1996,Huang1987}, provided that NASA COBE data are used and the magnitude of $\delta(p,\ x)$ is very small.

Analysis of NASA COBE data by Eq. (\ref{eq16}) is presented in Table \ref{tab2}.
\begin{table*}[htbp]
\caption{\label{tab2}Numerical proof of equivalences between Eqs. (\ref{eq2}) and (\ref{eq16}) is shown at typical observed values of NASA COBE data. Our result of analysis  by Eq. (\ref{eq16}) is as follows: $T=2.725008\pm 0.000026$ K, $|p-1|=8.09\times 10^{-5}$ and $\chi^2/{\rm NDF} = 45.01/41$. It should be noticed that we obtain the following values, $T=2.725016\pm 0.000008$ K and $\chi^2/{\rm NDF} = 45.10/42$, provided that the first term $F^{\rm (PD)}$ is only utilized.}
\begin{center}
\begin{tabular}{cccc}
\hline
$\nu/c$ [cm$^{-1}$] & 4.08 & 8.62 & 13.16\\
\hline
Eq. (\ref{eq2}) & 354.0636052 & 271.4238381 & 87.02961125\\
\hline
Eq. (\ref{eq16}) & 354.0636052 & 271.4238381 & 87.02961124\\
$F_1^{\rm (PD)}$ & 354.0594726 & 271.4209229 & 87.02868402\\
$F_2^{\rm (FC)}$ & 0.4.263569862$\times 10^{-3}$ & 2.920132007$\times 10^{-3}$ & 9.273229351$\times 10^{-4}$\\
$F_3^{\rm (FC)}$ & 1.310610486$\times 10^{-4}$ & 4.892780012$\times 10^{-6}$ & 1.016807344$\times 10^{-7}$\\
\hline
\end{tabular}
\end{center}
\end{table*}

\section{\label{sec4}Concluding remarks and discussion}
{\it C1)} We have shown that the ML function of Eq. (\ref{eq6}) is decomposed into two functions $e^x/p$ and $\delta(p,\ x)$: The various properties of the integral representation $\delta(p,\ x)$ are investigated. In particular, through the analysis of the COBE monopole data, it is shown that the magnitude of $\delta(p,\ x)$ is much smaller than that of the Planck distribution. See, Fig. \ref{fig3}a).

\noindent
{\it C2)} In our previous paper \cite{Biya2015}, we observed $|\mu| = |p-1|$ through the numerical analysis of the COBE monopole data by Eqs. (\ref{eq1}), (\ref{eq2}) and (\ref{eq4}). In the present paper, we have obtained Eq. (\ref{eq8}) i.e., $\mu= -\ln p$ in the analytic form. 

\noindent
{\it C3)} Combining the analytic relation of Eq. (\ref{eq8}) with the numerical results of Table \ref{tab2}, the following relation is obtained,
\begin{eqnarray}
  |\mu| = \ln(|p-1|+1).
\label{eq17}
\end{eqnarray}

\noindent
{\it D1)} Using the present study, the interesting result given by Eq. (\ref{eq17}) is achieved. In the near future, the effect of the fractional parameter on other fields (BE condensation) should be investigated. From Eqs. (\ref{eq2}) and (\ref{eq6}), we obtain the following formula,
\begin{eqnarray}
  g_{\nu}(p) = \frac 1{\Gamma(\nu)}\int_0^{\infty}\!\! \frac{x^{\nu-1}}{p^{-1}e^x-1} dx = \sum_{l=1}^{\infty} \frac{p^l}{l^{\nu}},\ (p<1),\quad
\label{eq18}
\end{eqnarray}
It is well-known that this formula is given in Refs. \cite{Pathria1996,Huang1987,Pethic2002}
\medskip\\

\section*{Acknowledgements}
 We would like to acknowledge Prof. M. Caputo for his kindness. Moreover, one of authors (M.B.) would like to thank the Department of Physics at Shinshu University for their hospitality.


\begin{thebibliography}{99}
\bibitem{Metzler2000} 
R. Metzler and J. Klafter, Phys. Rep. {\bf 339}, 1 (2000).

\bibitem{West2003} 
B. J. West, M. Bologna and P. Grigolini, \textit{Physics of Fractal Operators}, (Springer-Verlag, 2003).

\bibitem{Hilfer2000} 
R. Hilfer, J. Phys. Chem. {\bf 143}, 3914 (2000).

\bibitem{Barkai2001} 
E. Barkai,  Phys.\ Rev.\ E {\bf 63}, 046118 (2001).
 
\bibitem{Suzuki2001} 
N.~Suzuki and M.~Biyajima,  Phys.\ Rev.\ E {\bf 65}, 016123 (2001).

\bibitem{Biya2015} 
M. Biyajima, T. Mizoguchi and N. Suzuki, Physica {\bf A 440}, 129 (2015).

\bibitem{Caputo1967} 
M. Caputo, Geophys. J. R. Astr. Soc. {\bf 13}, 529 (1967); 
M.~Caputo and F.~Mainardi, Riv. Nuov. Cim. (Ser. II), {\bf 1}, 161 (1971); 
M.~Caputo, J. M. Carcione and M. A. B. Botelho, Fract. Calc. Appl. Anal. {\bf 18}, 3 (2015).

\bibitem{Ross1975} 
B. Ross (Ed.), \textit{Fractional Calculus and Its Applications}, Lecture Notes in Mathematics, vol. 457 (Springer-Verlag, New York, 1975).

\bibitem{Podlubny1999} 
I. Podlubny, \textit{Fractional Differential Equations} (Academic Press, San Diego, 1999).

\bibitem{Mainardi2001} 
F. Mainardi, Y. Luchiko and G. Panini, Frac. Calc. and Appl. Anal. {\bf 4}, 153 (2001).

\bibitem{Dzhe1993} 
M. M. Dzherbashian, \textit{ Harmonic analysis and Boundary Value Problems in the Complex Domain}, (Birkhauser, Basel, 1993).

\bibitem{Gore2002} 
R. Gorenflo, J. Loutchko and Y. Luchko, Frac. Calc. Appl. Anal. {\bf 5}, 491 (2002).

\bibitem{Gore2014} 
R. Gorenflo, A. A. Kilbas, F. Mainardi and S. V. Rogosin, \textit{Mittag-Leffler Functions, Related Topics and Applications }, (Springer, Heiderberg, 2014).

\bibitem{Mather1994} 
J.C. Mather et al., Astrophys. {\bf J. 420}, 439 (1994);
%
See also, 
D.J. Fixsen et al., Astrophys. {\bf J. 473}, 576 (1996); 
%
D.J. Fixsen and J.C. Mather, Astrophys. {\bf J. 581}, 817 (2002).

\bibitem{cobe2005} 
COBE/FIRAS CMB monopole spectrum, May 2005,\\
http://lambda.gsfc.nasa.gov/product/cobe/firas\_monopole\\
\_get.cfm ; 

\bibitem{Durrer2008} 
R.Durrer, \textit{The Cosmic Microwave Background}, (Cambridge University press, 2008).

\bibitem{Ertik2009} 
H. Ertik et al., Physica {\bf A 388}, 4573 (2009).

\bibitem{Biya2012} 
M. Biyajima and T. Mizoguchi, Phys. Lett. {\bf A 376} (2012) 3567;
See also, M.~Biyajima and T.~Mizoguchi, Astrophys. Space Sci. {\bf 350}, 317 (2014).

\bibitem{Komp1957} 
A. S. Kompaneets, Sov. Phy. JETP {\bf 4}, 730 (1957).

\bibitem{Weym1965} 
R. Weymann, Phys. Fluids {\bf 8}, 2112 (1965).

\bibitem{Planck1900a} 
M. Planck, {\it Ueber irreversible Strahlungsvorg\"ange}, Ann. d. Phys. {\bf 1}, 69 (1900). 

\bibitem{Planck1900b} 
M. Planck, Verh. Deutsch. Phys. Ges. {\bf 2}, 202 and 237 (1900).

\bibitem{Sommerfeld1956} 
A. Sommerfeld, \textit{Thermodynamics and Statistical Mechanics} (Academic Press, Yew York, 1956).

\bibitem{Tsallis2005} 
C. Tsallis, \textit{Introduction to Nonextensive Statistical Mechanics} (Springer Science+Business media LLC, 2009). 

\bibitem{Gradshteyn1965} 
I.S. Gradshteyn and I.M. Ryzhik, \textit{Table of Integrals, Series and Products}, (Academic Press, Yew York, 1965).

\bibitem{Pathria1996} 
R. K. Pathria, \textit{Statistical Mechanics}, (Second Ed., Elsevier, 1996).

\bibitem{Huang1987} 
K. Huang, \textit{Statistical Mechanics}, (Second ed., John Wiley \& Suns, Yew York, 1987). 

\bibitem{Pethic2002} 
C. J. Pethick and H. Smith, \textit{Bose-Einstein Condensation in Dilute Gases}, (Cambridge University Press, 2002).
\end{thebibliography}
\end{document}